\documentclass[elsart,floatfix,showpacs,noshowkeys,epsfig,graphics]{revtex4}%
\usepackage{graphicx}
\usepackage{amsmath}
\usepackage{amsfonts}
\usepackage{amssymb}

\newcommand{\newc}{\newcommand}
\newc{\beq}    {\begin{equation}}
\newc{\eeq}    {\end{equation}}

\newc{\beqa}    {\begin{eqnarray}}
\newc{\eeqa}    {\end{eqnarray}}
\newc{\bs}    {\section}
\newc{\no}    {\\ \nonumber}

\begin{document}
\title{Entanglement of  thermal scalar fields in a compact space}
\author{Jae-Weon Lee}
\email{scikid@kias.re.kr}
\author{Jaewan Kim}
\affiliation{School of Computational Sciences,
             Korea Institute for Advanced Study,
             207-43 Cheongnyangni 2-dong, Dongdaemun-gu, Seoul 130-012, Korea}
\author{Taeseung Choi}
\affiliation{Department of Physics,
             Korea university,Anam-dong Seongbuk-Gu, Seoul, 136-701, Korea}
\date{\today}
\begin{abstract}
Using  the thermal  Green's function approach we propose a general method to investigate entanglement  of the vacuum state or
thermal ground states in an arbitrary dimensional space-time.
As an application we show quantum separability of
the massive thermal scalar field vacuum in the 1+1 dimensional cylindrical space-time. Separability is demonstrated
using the positive partial transpose criterion for effective
two-mode Gaussian states of collective operators. In this case, for all mass and temperature values
entanglement is absent between the collective operators.
\end{abstract}

\maketitle

Entanglement is now treated as a physical quantity,
as well as valuable resource allowing various quantum information processing.
This  brings a renewal of interest in studying the non-locality
and entanglement in many particle systems such as Bose Einstein condensations~\cite{bec},
fermion systems~\cite{Vedral03,oh}, and superconductors~\cite{ohsc}.
 Vedral~\cite{Vedral03}
studied the entanglement in many body systems at zero temperature
using the second quantization formalism. Following his works, Oh
and Kim~\cite{oh} studied the entanglement of  two electron spins
in a free electron gas, superconductivity~\cite{ohsc} and the Kondo model~\cite{kondo} at finite
temperature using thermal Green's function methods.
Recently, in the context of AdS/CFT duality~\cite{Maldacena},
entanglement
entropy~~\cite{ryu:181602,emparan-2006-0606,PhysRevD.34.373} is used
 in studying the black hole entropy~\cite{PhysRevD.34.373,solodukhin:201601}.
It was also  suggested~\cite{entanglementDE,ourcoincidence}  that entanglement is an origin of
mysterious dark energy.
However, since entanglement entropy is a good measure of entanglement only for pure states,
 we need a reliable method to calculate entanglement for more general states.
Thus, studying entanglement of quantum fields vacuum or thermal ground states is of fundamental
importance~\cite{reznik}
in various physical fields.
In ~\cite{my}, the temperature Green's function approach
was used to
investigate the quantum entanglement of two non-interacting
 spin 1 boson particles in  thermal
equilibrium.
It is also well known that, in algebraic quantum field theory, due to the Reeh-Schlieder theorem~\cite{reeh}
 the vacuum
for  quantum fields violate Bell inequality and has quantum
nonlocality~\cite{werner,narnhofer1}.
In Ref. ~\cite{narnhofer1,narnhofer2} entanglements of  Bose and Fermi quasi-free (gaussian) states
are extensively studied, while spatial entanglement of
free thermal bosonic fields
was studied by  post-selecting certain momenta in Ref. ~\cite{heaney-2006}.
 On the other hand,  for decades the nature of the  vacuum, concerning the
Casimir energy~\cite{casimir,casimir2}, in a bounded space has been
extensively studied using the Hadamard Green's function
method~\cite{dewitt}.
In Ref. ~\cite{our} we showed that this Hadamard Green's function is also useful
to study entanglement of vacuum and used the method to investigate entanglement of massless scalar field
vacuum with
a Dirichlet boundary.
Entanglement of the scalar field could be experimentally tested by the scheme with trapped ions~\cite{reznik,reznik:042104}
  or Bose-Einstein condensates~\cite{Kaszlikowski}.
The entanglement measures such as the purity~\cite{purity} or the
negativity has been also suggested~\cite{negativity} for two-mode
Gaussian states.
Since
the scalar fields are continuous variables we need a separability
criterion for continuous variables such as Gaussian states.
 To utilize positive partial transpose (PPT) criterion for Gaussian mode, we  averaged the fields over two
tiny boxes to make the vacuum state an effective two-mode Gaussian
state.

In this letter, we first generalize the methods proposed in Ref. ~\cite{our}
to arbitrary space-time dimension and use this   Green's function
approach  to investigate the  quantum entanglement of
the vacuum
for
a massive thermal scalar field in the 1+1 dimensional cylindrical space-time.
Studying the nature  of quantum fields in a compact space is especially important in the string physics.
Our methods could be used to study entanglement of continuum limit of harmonic chains
in various situations.

 We start by introducing  massive real Klein-Gordon  fields, which are described by a Lagrangian
 density,
 \beq
 \label{L}
 \mathcal{L} (\vec{x},t)=\frac{1}{2}( \partial_\alpha\phi
\partial^\alpha \phi +m^2 \phi^2),
\eeq
 in  $(D+1)$ dimensional space-time.
The equal time commutation relations of the scalar field are
\beqa
\label{commutation}
 [ \phi(\vec{x},t),\phi(\vec{x}',t)]&=&0 , [ \pi(\vec{x},t),\pi(\vec{x}',t)]=0, \no
 [ \phi(\vec{x},t), \pi(\vec{x}',t) ] &=& i \delta(\vec{x}-\vec{x}'),
 \eeqa
where $\pi(\vec{x},t)\equiv \partial_t \phi(\vec{x},t)$ is a
momentum operator for $\phi(\vec{x},t)$. Consider a vector
of the field and its momentum operator at two points $\vec{x}$ and
$\vec{x}'$
 at a given time $t$, i. e., $\xi\equiv (\phi(\vec{x},t),\pi(\vec{x},t),\phi(\vec{x}',t),\pi(\vec{x}',t))$.
The variance matrix for the vacuum or the thermal ground state of the scalar field  ($|0\rangle$)
 is defined as
 \beq
 V_{\alpha \beta}\equiv \frac{1}{2}\langle 0|\{\triangle \xi_\alpha,  \triangle \xi_\beta  \}|0\rangle,
 \eeq
 where $\{A,B\}=AB+BA$  and $\triangle \xi_\alpha\equiv \xi_\alpha-\langle 0| \xi_\alpha|0\rangle$
 with $\langle 0| \xi_\alpha|0\rangle=0$ in this paper.
 Following Ref. \cite{our}, one can obtain  the variance matrix
\begin{eqnarray}
V =            \begin{bmatrix}
            a & 0 & c  &0 \\
            0 & b & 0 &d \\
            c & 0 & a'  &0 \\
            0 & d & 0 & b'
           \end{bmatrix},
\end{eqnarray}
where $a=\langle0|\{\phi(\vec{x},t),\phi(\vec{x},t)\}|0\rangle/2$,
$b=\langle
0|\{\pi(\vec{x},t),\pi(\vec{x},t)\}|0\rangle/2, ~a'=\langle0|\{\phi(\vec{x'},t),\phi(\vec{x'},t)\}|0\rangle/2,$
and so on.
 Usually  $a$ and $a'$ terms have  a divergence which can be removed
 by  subtracting a
free-space Green's function $G_0$ from a Green's function $G$ called the Hadamard's
elementary function ~\cite{birrell}.  Thus we can define a regularized  Green's
function; $ G_R\equiv G-G_0$ which is not divergent.

In Ref. \cite{our} we showed that,
once we know the Green's function we can calculate the components of the variance matrix $V$ from it, i.e.,
 \beqa
 \label{cd}
c(\vec{x},\vec{x}')&=&\frac{1}{2} \langle
0|\{\phi(\vec{x},t),\phi(\vec{x}',t)\}|0\rangle =\lim_{t\rightarrow
t'}\frac{1}{2} G_R(\vec{x},t;\vec{x}',t'),\no
d(\vec{x},\vec{x}')&=&\frac{1}{2}
\langle 0|\{\partial_t\phi(\vec{x},t),\partial_t\phi(\vec{x}',t)\}|0\rangle =
\lim_{t\rightarrow t'} \partial_t \partial_{t'}
\frac{1}{2}G_R(\vec{x},t;\vec{x}',t').
 \eeqa
Then, $ a= \lim_{\vec{x}' \rightarrow \vec{x}}
c(\vec{x},\vec{x}')$, $ b= \lim_{\vec{x}'\rightarrow \vec{x}}$
$ d(\vec{x},\vec{x}')$ and so on.

\begin{figure}[htbp]
\includegraphics[width=0.4\textwidth]{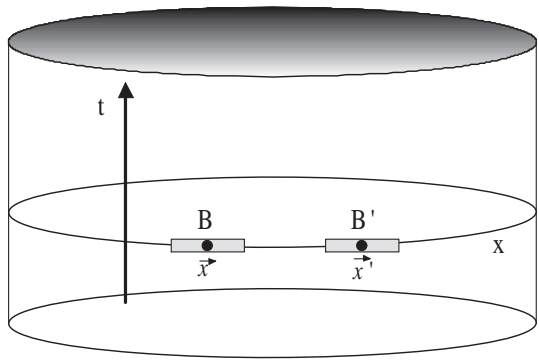}
\caption{Cylindrical 1+1 dimensional space time with spatial circumstance $R$.
 we average  fields over two tiny boxes $B$ and $B'$  centered at $x$ and $x'$, respectively.
\label{ccorrelation} }
\end{figure}
 Since `if and only
if ' separability test for infinite-mode states are unknown, we
need to
 reduce the infinite-mode states to effective two-mode Gaussian states by defining collective
operators. To define collective operators, we follow
the approach in Ref. ~\cite{kofler:052107}, i.e., at a given time $t$
we spatially average the field operator over  two tiny spatial boxes $B$ and $B'$  centered at $\vec{x}_0$ and $\vec{x'}_0$,
respectively (See Fig. 1).
Defining collective operators in this manner is physically reasonable, because
in a real situation probes always have  finite spatial resolution.
We do not average but integrate the momentum operators within the box, since momentum is additive.
( However, averaging the momentum gives the same final results in this work.)
  Thus the collective operators are
  \beqa
\Xi&\equiv& \left(\Phi(\vec{x},t),\Pi(\vec{x},t),\Phi(\vec{x'},t),\Pi(\vec{x'},t)\right)\no
&=&\left(\frac{1}{L^D}\int_Bd^D \vec{y} \phi(\vec{x}+\vec{y},t),
  \int_Bd^D \vec{y} \pi(\vec{x}+\vec{y},t),
  \frac{1}{L^D}\int_{B'}d^D\vec{y'} \phi(\vec{x'}+\vec{y'},t),
  \int_{B'}d^D\vec{y'} \pi(\vec{x'}+\vec{y'},t)\right).
  \eeqa

Here $\int_B d^D\vec{y} f(\vec{x}+\vec{y})$ denotes an integration of $f(\vec{x})$ over a box $B$
 centered at $\vec{x}$ with volume $L^D$.
  Then, the commutation relations in Eq. (\ref{commutation})
  reduce to the  canonical commutation relations for the collective operators  ~\cite{kofler:052107}
\beq
[{\Xi}_\alpha,{\Xi}_\beta]=i\Omega_{\alpha\beta},
\eeq
where
\begin{eqnarray}
\Omega =            \begin{bmatrix}
            J & 0  \\
            0 & J \\
           \end{bmatrix}\,,
           J =            \begin{bmatrix}
            0 & 1  \\
            -1 & 0 \\
           \end{bmatrix}\,.
\end{eqnarray}
Similarly we define a variance matrix for the collective operators
$\tilde{V}_{\alpha \beta}\equiv \frac{1}{2}\langle 0|\{ {\Xi}_\alpha,  {\Xi}_\beta  \}|0\rangle.$

Now we discuss separability of the vacuum.
Using the fundamental theorem of calculus~\cite{calculus},
one can calculate the  variance matrix $\tilde{V}$ for the two-mode Gaussian states
with those of $V$ in the limit $L\rightarrow 0$ (but still non-zero).
For example,
\beqa
\tilde{V}_{13}&=&\lim_{L\rightarrow 0}\frac{1}{2}\langle 0|\{ \Phi(\vec{x},t),  \Phi(\vec{x'},t)\}|0\rangle \no
&=&\lim_{L\rightarrow 0} \frac{1}{2L^{2D}}  \int_Bd^D\vec{y}\int_{B'}d^D\vec{y}'\langle 0|\{ \phi(\vec{x}+\vec{y},t),
 \phi(\vec{x'}+\vec{y}',t)\}|0\rangle \no
&=&\frac{1}{2}\langle 0|\{ \phi(\vec{x},t), \phi(\vec{x'},t)\}|0\rangle
=V_{13}=c.
\eeqa
This is always possible when  the integrand of the second line
 is continuous.
 Similarly,
\beqa
\tilde{V}_{24}&=&
\lim_{L\rightarrow 0}  L^{2D}\frac{1}{2L^{2D}}  \int_Bd^D\vec{y}\int_{B'}d^D\vec{y}'\langle 0|\{ \pi(\vec{x}+\vec{y},t),
 \pi(\vec{x'}+\vec{y}',t)\}|0\rangle =L^{2D} V_{24}=L^{2D} d.
\eeqa
Hence,
\begin{eqnarray}
\label{Vnew}
\tilde{V} =
             \begin{bmatrix}
            a & 0 & c  &0 \\
            0 &L^{2D} b & 0 & L^{2D} d \\
            c & 0 & a'  &0 \\
            0 & L^{2D} d & 0 & L^{2D} b'
           \end{bmatrix}\,
\equiv
              \begin{bmatrix}
            A & G  \\
            G^T & B  \\
           \end{bmatrix}.
\end{eqnarray}
For $L$ infinitesimally small the above equation
gives an exact  result, while for $L \ll 1$ it provides us a good approximation.
The separability  criterion we use in this paper is  PPT criterion~\cite{peres,ppt}
for two-mode Gaussian states~\cite{simon}
which is equivalent to
\beq
F\equiv \tilde{\Sigma}-(\frac{1}{4}+4 det \tilde{V})\leq 0,
\label{F}
\eeq
where $\tilde{\Sigma}=det A +det B-2det G$.
By inserting the components of $\tilde{V}$ into Eq. (\ref{F}) we obtain
\beq
F=-\frac{1}{4}+L^{2D}[ab+a'b'-2cd]-4L^{4D}[aa'bb'-aa'd^2-c^2bb'+c^2d^2].
\eeq
Now the quantum separability can be tested for an arbitrary space-time
using this quantity once the Hadamard's elementary function is known.
Furthermore, if $a=a'$ and $b=b'$ one can immediately obtain an entanglement measure called the negativity
using Eq. (9) in Ref. ~\cite{PhysRevA.66.030301};
\beq
E(\rho)=max\left\{0,\frac{1}{L^{2D}(a-|c|)(b-|d|)}-1\right\}.
\eeq
Thus, we can easily determine how much a given state has entanglement, once we know the Green's function
at least for the collective operators. This is very useful results since the Hadamard's function is well known
for many situations.

As an application, we test the quantum separability of the massive thermal scalar fields vacuum
in 1+1 dimensional cylindrical space-time using the methods described above.
This model is related to one-dimensional harmonic chain systems with periodic boundary condition
\cite{Cramer:2005mx,Plenio:2004he,kofler:052107}.
First let us choose two
points $x$ and $x'$ (see Fig. 1). Using the
method of images~\cite{image} one can obtain the thermal Hadamard Green's function for this field
~\cite{birrell}
\beq
G = \frac{-1}{4 \pi} ln\left(16 \left[ sin(\pi (\Delta u+i \beta m)/R)sin(\pi (\Delta v+i \beta m)/R)\right]^2\right),
 \eeq
where $u \equiv -x + t $,
$ v \equiv x  + t$
are the light-cone coordinates, and $\Delta u=-(x'-x)+(t'-t)\equiv -\Delta x +\Delta t$.
 This Green's function diverges as $R \rightarrow \infty$;
\beq
\lim_{R\rightarrow \infty} G\sim -\left( \frac{\ln (\frac{1}{R})}{\pi } \right)  - \frac{\ln (16\,
      {\left( \pi{\Delta t}  - \pi{\Delta x}  + i \pi\,m\, \,\beta  \right) }^2\,
      {\left( \pi{\Delta t}  + \pi {\Delta x} + i \pi\,m\, \,\beta  \right) }^2)}{4\,\pi }\equiv G_0.
\eeq
Interestingly, the Green's function is a function of $m\beta\equiv M$.
As mentioned above by subtracting  $G_0$
from $G$ one can define a regularized  Green's
function $ G_R\equiv G-G_0$, from which one can obtain
\beqa
& &(a,b,c,d)= \no
& &\left( \frac{1}{2\,\pi }{\ln \left(\frac{\pi m\, \,\beta \,csch(\frac{\pi m\, \,\beta }{R})}{R}\right)},\frac{-\pi }{6\,R^2},
 \frac{1}{8\,\pi }{\ln \left(\frac{16\,{\pi }^4\,{\left( {{\Delta x}}^2 + m^2\,{\beta }^2 \right) }^2}{4R^4\,
{\left( \cos (\frac{2\,{\Delta x}\,\pi }{R}) -
  \cosh (\frac{2\,m\,\pi \,\beta }{R}) \right) }^2}\right)},
  \frac{1}{2\,\pi }\left({{\Delta x}}^{-2} - \frac{{\pi }^2\,{\csc^2 (\frac{{\Delta x}\,\pi }{R})}}{R^2}\right)\right)
\eeqa
using Eq. (\ref{cd}) after some tedious calculation.
By inserting these components into Eq. (\ref{F}) we obtain
\beq
 F=-\frac{1}{4} +L^2 f_2(\Delta x, M) +L^4 f_4(\Delta x, M) ,
\eeq
where
\beq
f_2(\Delta x, M) \equiv -\frac{ \left( -1 + 2\,{\pi }^2\,{{\Delta
x}}^2 + \cos (2\,\pi \,{\Delta x}) \right) \,
       {\csc^2 (\pi \,{\Delta x})}\,\ln \left(\frac{{\left( \cos (2\,\pi \,{\Delta x}) -
              \cosh (2\,M\,\pi ) \right) }^2}{4\,{\pi }^4\,{\left( M^2 + {{\Delta x}}^2 \right) }^2}\right)
        }{16\,{\pi }^2\,{{\Delta x}}^2} - \frac{\ln (M\,\pi \,{csch}(M\,\pi ))}{6}
\eeq
and
\beq
f_4(\Delta x, M) \equiv
\frac{-1}{576\,{\pi }^4}{\left(  9{\left( {{\Delta x}}^{-2} - {\pi }^2\,{\csc (\pi \,{\Delta x})}^2
           \right) }^2 -{\pi }^4\right) \,
       \left( \left[\ln \left(\frac{{\left( \cos (2\,\pi \,{\Delta x}) -
               \cosh (2\,M\,\pi ) \right) }^2}{4\,{\pi }^4\,{\left( M^2 + {{\Delta x}}^2 \right) }^2}\right)\right]^2 -
      16\,[\ln (M\,\pi \,{csch}(M\,\pi ))]^2 \right) }.
\eeq
Here we have set $R=1$, hence from now on all length scales are in units of $R$.
 \begin{figure}[htbp]
\includegraphics[width=0.4\textwidth]{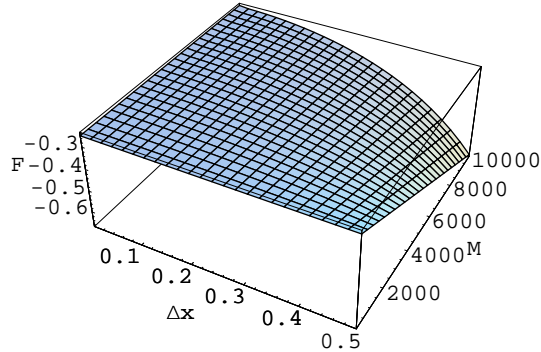}
\caption{$F$ as a function of $\Delta x$ and $M$ for L=0.01.
 \label{Fig2} }
\end{figure}

Our numerical study shows that $f_1$ has a
maximum $f_1(1/2,0)=0.134$ and  $f_2$ has a
maximum $f_2(1/2,1.107)=0.0164$.
 Hence, one can see that
\beq
F<-\frac{1}{4} +0.134 ~ L^2 +0.0164 L^4 <0
\eeq
for $0<L<1/2$.
Therefore, the effective two-mode scalar field vacuum with the periodic boundary is PPT and, hence, separable,
for all $M$ and $\Delta x$.
Fig. 2 shows $F$ as a function of $\Delta x$ and $M$ for $L=0.01$.

Our results  implies that changing mass, temperature, and periodic boundary condition
does not induce additional entanglement to the possible entanglement of the
 zero temperature free space scalar fields vacuum.
Our results also indicates that the entanglement of generic vacuum resides in the
free space Green's function $G_0$ which usually gives rise to infinite quantities.
Our approach
provides a new and generic method  to
investigate entanglement of the quantum fields vacuum or thermal ground states in a various situation
 using the Green's function method. Since the Hadamard's functions are known for many cases
our approach can be useful.

\vskip 15pt
\paragraph*{Acknowledgments.--}
\indent
Authors are thankful to Jinhyoung Lee for helpful discussions.
J. Lee and
J. Kim
was supported by the Korea Ministry of Information and Communication
with the``Next Generation Security Project".
 T. Choi was supported by the SRC/ERC program of MOST/KOSEF
(R11-2000-071), the Korea Research Foundation Grant
(KRF-2005-070-C00055), the SK Fund, and the KIAS.


\end{document}